# Unraveling the Generation Mechanism of a Mid-Latitude Plasma Blob and the Evidence of Its Rare Interaction with a MSTID Phase Front

D. Patgiri[1], R. Rathi[2], V. Yadav[3], D. Chakrabarty[4], M. V. Sunil Krishna[1,5], S. Kannaujiya[6], S. Sunda[7], Y. Otsuka[8], A. Shinbori[8], M. Nishioka[9], S. Perwitasari[9], P. Pavan Chaitanya[10], S. Sarkhel[1,5,*]

[*]Sumanta Sarkhel, Department of Physics, Indian Institute of Technology Roorkee, Roorkee - 247667, Uttarakhand, India (sarkhel@ph.iitr.ac.in)

[1]Department of Physics, Indian Institute of Technology Roorkee, Roorkee – 247667, Uttarakhand, India

[2]School of Physics and Astronomy, Lancaster University, Lancaster, United Kingdom

[3]Aryabhatta Research Institute of Observational Sciences, Nainital – 263001, Uttarakhand, India

[4]Physical Research Laboratory, Space and Atmospheric Sciences Division, Ahmedabad, India

[5]Centre for Space Science and Technology, Indian Institute of Technology Roorkee, Roorkee –247667, Uttarakhand, India

[6]Indian Institute of Remote Sensing, ISRO, Dehradun - 248001, India

[7]GNSS Research Center for Ionospheric Studies & Navigation Application, Airports Authority of India, Ahmedabad, India

[8]Institute for Space-Earth Environmental Research, Nagoya University, Nagoya, Japan

[9]National Institute of Information and Communications Technology (NICT), Japan

[10]National Atmospheric Research Laboratory, Gadanki, India.

## Abstract

We report observations of two distinct nighttime F-region irregularities, plasma blob (localized density enhancement) and medium-scale traveling ionospheric disturbance (MSTID), in O($^1$D) 630.0 nm all-sky airglow images from Hanle (32.7°N, 78.9°E; Mlat~24.1°N), Ladakh, India, during the geomagnetically quiet (Ap=6) night of 06 July 2021. Global vertical total electron content (VTEC) maps revealed that the plasma blob developed beyond the southern edge of imager's field-of-view before appearing in airglow images and propagated predominantly westward, as confirmed from both the airglow and VTEC datasets. The existence of the plasma blob and MSTID outside the imager's field-of-view was further confirmed by temporal VTEC fluctuations recorded by multiple GNSS receivers. Additionally, FORMOSAT-7/COSMIC-2 signal-to-noise ratio and ICON/MIGHTI wind profiles indicated the presence of sporadic-E ($E_S$) layers at E-region near both the plasma blob and the MSTID. We propose that polarization electric field associated with either MSTID or $E_S$-layers mapped along magnetic field lines to lower latitudes, driving upward plasma transport from F-peak region through vertical uplift of the F-layer. This F-layer uplift was confirmed by simultaneous in-situ $O^+$/$H^+$ density enhancements/reductions at LEO altitudes measured by FORMOSAT-7/COSMIC-2. Upward-transported plasma experienced reduced chemical loss at higher altitudes, producing localized VTEC enhancements (plasma blob). The plasma subsequently diffused along magnetic field lines to higher/lower latitudes/altitudes (~250 km), entering imager's field-of-view, where enhanced dissociative recombination of $O_2^+$ produced high intensity airglow region. Interestingly, interaction between the plasma blob and MSTID's plasma-depleted front caused gradual decay and bifurcation of the front due to plasma influx from the high-density blob region.

**Key Points:**

1. The polarization electric field of plasma-depleted front of the MSTID was responsible for the generation of plasma blob.
2. The ascent of F-layer led to enhancement/reduction of $O^+$/$H^+$ densities measured by FORMOSAT-7/COSMIC-2 at LEO altitudes.
3. Plasma blob interacted with MSTID's plasma-depleted phase front, causing gradual decay and bifurcation of the front.

## 1. Introduction

Medium-scale traveling ionospheric disturbances (MSTIDs) at midlatitudes are among the most prominent and widely studied ionospheric irregularities and they have been investigated using a variety of observational techniques (Ding et al., 2011; Otsuka et al., 2007, 2011). MSTIDs are a special class of TIDs characterized by quasi-periodic variations in ionospheric plasma density, with horizontal wavelengths up to a few hundred kilometers and horizontal phase speeds up to a few hundred meters per second (Ding et al., 2011; Otsuka et al., 2011; Shiokawa et al., 2003a). MSTIDs can be classified into daytime, nighttime, dawn-time, and dusk-time categories based on their period of occurrence (Ding et al., 2011; Otsuka et al., 2011, 2021, Shiokawa et al., 2003a). In particular, nighttime MSTIDs at mid-latitudes are driven by the well-established electrodynamic instability mechanism known as the Perkins instability (Hamza, 1999; Perkins, 1973). Coupled electrodynamics between the E- and F-regions serves as a feedback mechanism to overcome the instability's inherently low growth rate (Cosgrove and Tsunoda, 2004; Narayanan et al., 2018; Otsuka et al., 2007; Patgiri et al., 2024c; Tsunoda. 2006; Yokoyama et al., 2009). Generally, the phase fronts of nighttime MSTIDs are preferentially aligned along the northwest-southeast direction and propagate southwestward, although cases contrasting with these characteristics have been reported in the literature (Ding et al., 2011; Narayanan et al., 2018; Otsuka et al., 2011; Patgiri et al., 2024a; Rathi et al., 2021, 2024, 2025; Yadav et al., 2021a).

In contrast to MSTIDs, equatorial plasma bubbles (EPBs) are equatorial and low-latitude F-region ionospheric irregularities that occur predominantly after sunset and are characterized by strong plasma density depletions relative to the ambient ionosphere (Bhattacharyya, 2022; Burke et al., 1980; Tsunoda et al., 1982). Historically, coherent radar backscatter observations of ionospheric plumes have revealed the upwelling of EPBs, which are driven by the Rayleigh-Taylor instability (Kelley, 2009; Sekar et al., 2008; Woodman and LaHoz, 1976). While the Rayleigh-Taylor mechanism exhibits linear instability exclusively within the bottom-side F region, its nonlinear evolution causes these perturbations to rise upward and penetrate into the topside ionosphere (Sekar and Chakrabarty, 2011; Huba et al., 2009; Keskinen et al., 2003; Scannapieco and Ossakow, 1976; Zalesak et al., 1982).

Another prominent F-region phenomenon involves localized plasma density enhancements, commonly referred to as plasma blobs, which have attracted significant interest from the space science community since their discovery by Oya et al. (1986) and Watanabe and Oya (1986). Although no independent generation mechanism for plasma blobs has been definitively established, most studies report their formation as closely related to EPBs (Anderson & Haerendel, 1979; Haaser et al., 2012; Huang et al., 2014; Krall et al., 2009, 2010; Le et al., 2003; Martinis et al., 2009; Park et al., 2003, 2008; Yokoyama et al., 2007). 2D numerical simulations predicted plasma blobs alongside EPBs (Sekar et al., 2001) and their existence was confirmed (Sekar et al., 2004) by narrow spectral band, narrow field-of-view airglow photometry. The idea of a bubble-blob connection is mainly based on observations that plasma bubbles and blobs often occur along the same magnetic meridian and exhibit similar spatial distribution patterns. Thus far, three plasma blob generation mechanisms associated with EPBs have been identified. First, polarization electric fields within low-latitude bubbles map along geomagnetic field lines to EIA crests, transporting dense plasma upward to form blobs (Le et al., 2003). Second, EPBs initiating below the F2 peak develop moderate equatorial density enhancements relative to ambient levels as their apexes rise above the peak (Anderson and Haerendel, 1979; Haaser et al., 2012). Third, simulations show meridional winds redistributing plasma within EPB flux tubes to create localized density peaks (blobs) at flux tube ends (Krall et al., 2009, 2010; Huang et al., 2014).

It is to be noted that plasma blobs have also been observed without accompanying EPBs as well as during MSTID events (Choi et al., 2012; Kil and Paxton, 2017; Kil et al., 2011, 2019; Miller et al., 2014). These observations indicate possible alternative generation mechanisms, although their precise electrodynamics remains undefined. Using multi-instrument observations, this study reports the simultaneous occurrence of MSTIDs and plasma blobs on the geomagnetically quiet night of 06 July 2021 over the low-to-midlatitude transition region. It demonstrates that MSTID-associated electrodynamics and the plasma blob formation are causally linked. The primary dataset for this study, O($^1$D) 630.0 nm airglow images recorded by an all-sky imager located at Hanle, Ladakh, in the Himalayan region, a proven vantage point for observing diverse F-region ionospheric irregularities (Bhardwaj et al., 2023; Chakrabarti et al., 2024, 2025; Patgiri et al., 2024a, 2024b, 2024c; Rathi et al., 2021, 2022, 2024, 2025; Sivakandan et al., 2020, 2021; Yadav et al., 2021a, 2021b). However, plasma blob detections with this imager have not been reported previously. Additionally, this study also demonstrates between two distinct plasma irregularity structures: MSTID and plasma blob.

## 2. Instrument and data analyses

In the present study, we used $O(^1D)$ 630.0 nm unwarped images of an all-sky airglow imager installed at Hanle (32.7°N, 78.9°E; Mlat. ~ 24.1°N), Ladakh, India, on the geomagnetically quiet (Ap=6) night of 06 July 2021. The images captured on the previous quiet (Ap=7) night, 05 July 2021, were also utilized for comparing the novel observation of 06 July 2021. Readers can find a comprehensive description of the imager, as well as geospatial calibration and image-processing methodology, in Mondal et al. (2019).

In order to examine the temporal fluctuation of VTEC associated with the observed plasma blob and MSTID on 06 July 2021, we have utilized VTEC data from five Global Navigation Satellite System (GNSS) receivers, viz., GNFC (30.5°N, 78.1°E), BAGD (27.2°N, 88.8°E), LUCK (27°N, 80.7°E), KUGE (27.6°N, 85.5°E), and SNDL (27.4°N, 85.8°E). The receiver-independent exchange (RINEX) observation and navigation files were processed through software developed by Seemala (2022) to obtain the VTEC values. While utilizing the VTEC values from each receiver, an elevation angle cut-off of 35° is kept to avoid the multi-path effect.

Furthermore, the international GNSS service (IGS) global VTEC maps were used to examine the evolution of the plasma blob, as reflected in the spatiotemporal variations in VTEC data. These VTEC maps are generated by Universitat Politècnica de Catalunya at rapid (1-day) latencies through the regional VTEC modelling from the TOMographic IOnosphere (TOMION-v1) model software (Liu et al., 2021). A kriging interpolation method is adopted for the regional VTEC values to obtain 15-minute cadence maps (Orús et al., 2005). Then, bivariate interpolation in latitude, longitude, and linear time interpolation is done while providing the VTEC values in a grid of 2.5° (latitude) × 5° (longitude) at arbitrary positions and times (Schaer et al. 1998). In the present study, global VTEC data for the event night (06 July 2021) are utilized and compared with those from a control quiet night (05 July 2021).

Besides VTEC measurements, in-situ ion densities ($O^+$, $H^+$, and total) from the ion velocity meter (IVM) onboard FORMOSAT-7/COSMIC-2 (Chou et al., 2021; Wu et al., 2022) satellite were also utilized to examine the variation of ion densities inside the plasma blob. In order to examine the background F-layer electron density prior to the formation of the plasma blob, electron density profiles of FORMOSAT-7/COSMIC-2 (Lin et al., 2020) were used.

The rate of TEC index (ROTI) maps calculated as the standard deviation of the rate of TEC (differential of TEC within 30s intervals) are used (Pi et al., 1997) to check the presence of any ionospheric irregularities at low-latitudes. We used ROTI data that were mapped onto geographical coordinates, assuming an ionospheric altitude of 300 km with spatial and temporal resolutions of 0.5° (longitude) × 0.5° (latitude) and 5 minutes, respectively (Sori et al., 2021). Additionally, in order to check the spread-F conditions over the low-latitude regions, ionograms recorded by digisonde located at Gadanki (13.5°N, 79.2°E; Mlat. ~6.5°N) were also utilized.

It is noteworthy that F-region irregularities, such as MSTIDs, are closely linked to spatial inhomogeneities in the Sporadic-E ($E_S$) layer, as the two regions are electrodynamically coupled (Otsuka et al., 2007; Patgiri et al., 2024c). $E_S$-layers are often formed at the convergence node of vertical shear in zonal neutral wind (Axford, 1963). In order to investigate the presence of $E_S$-layers on 06 July 2021, the cardinal wind data (level 2.2) of the green line measured by the Michelson Interferometer for Global High-Resolution Thermospheric Imaging (MIGHTI) onboard the Ionospheric Connection Explorer (ICON) spacecraft were used (Englert et al., 2017; Harding et al., 2021). In the present study, the "1 sample" (1-sigma estimation) uncertainty associated with the wind data has been used as the error bar for each wind profile. In addition, since signal-to-noise (SNR) of the $L_1$ band is prominently affected by the gradient in plasma density caused by $E_S$-layers (Arras & Wickert, 2018; Yamazaki et al., 2022), the FORMOSAT-7/COSMIC-2 SNR radio occultation (RO) profiles of the $L_1$ band were further used to examine the presence of $E_S$-layers.

## 3. Results

Leveraging coordinated ground and space-based observations, we have examined the generation and evolution of a plasma blob concurrently detected with MSTID structure in $O(^1D)$ 630.0 nm all-sky airglow images from Hanle, Ladakh, India, on the geomagnetically quiet night of 06 July 2021. The plasma blob and MSTID appeared in the airglow images as a localized region of enhanced airglow intensity and northwest-southeast aligned phase fronts of low airglow intensity, respectively. Interestingly, on the previous geomagnetically quiet night (05 July 2021), no plasma blob was detected in the $O(^1D)$ 630.0 nm airglow images. Only a few MSTID phase fronts exhibiting very faint intensity were detected in the airglow images. For the reader's convenience, a Movie 1 showing the sequence of airglow images captured on both nights has been provided in the supplementary information. The detailed analyses and the interesting results are described in the following sections.

### 3.1 Simultaneous detection of plasma blob and MSTID fronts in all-sky airglow images

Figures 1a-o represent the sequence of O($^1$D) 630.0 nm airglow images recorded by the Hanle imager on 06 July 2021. The colored stars and dashed-dotted lines in these images represent the location of the Hanle imager and the geomagnetic field over Hanle, respectively. Initially, at 15:08:09 UT, two plasma-depleted fronts of the MSTID were first observed as dark fronts (or reduced airglow intensity regions) in the airglow images, identified as '$d_1$' and '$d_2$'. A localized region of enhanced airglow intensity (hereafter referred to as the plasma blob, '$b_1$') began to appear between '$d_1$' and '$d_2$' near the southernmost part of the airglow imager's field-of-view (FOV) around 15:45:40 UT (Figure 1c). After this time, the plasma blob evolved while propagating westward with an average velocity of 45 ± 8 m/s (Figures 1c-1i). The velocity of the plasma blob was calculated by tracking the maximum airglow intensity associated with the plasma blob along east-west direction (Figure S1 in the supplementary information). Additionally, the MSTID phase fronts, '$d_1$' and '$d_2$', propagated southwestward with an average velocity of 53 ± 9 m/s. The procedure for the calculation of the horizontal velocity of MSTID is described in Patgiri et al. (2024a). Interestingly, the plasma blob and the MSTID dark front, '$d_1$', engaged in a dynamic interaction. After the interaction, the central part of '$d_1$' started distorting as indicated by blue colored arrows in Figures 1g-1l. From 16:35:41 UT, the front seemed to be appearing as two distinct fronts, '$d_1^a$' and '$d_1^b$' (Figures 1h-1l). In order to confirm the uniqueness of this event, we also analyzed the airglow observations from the preceding geomagnetically quiet night (05 July 2021). However, no plasma blob activity was evident in those images, which are provided in the supplementary file (S1).

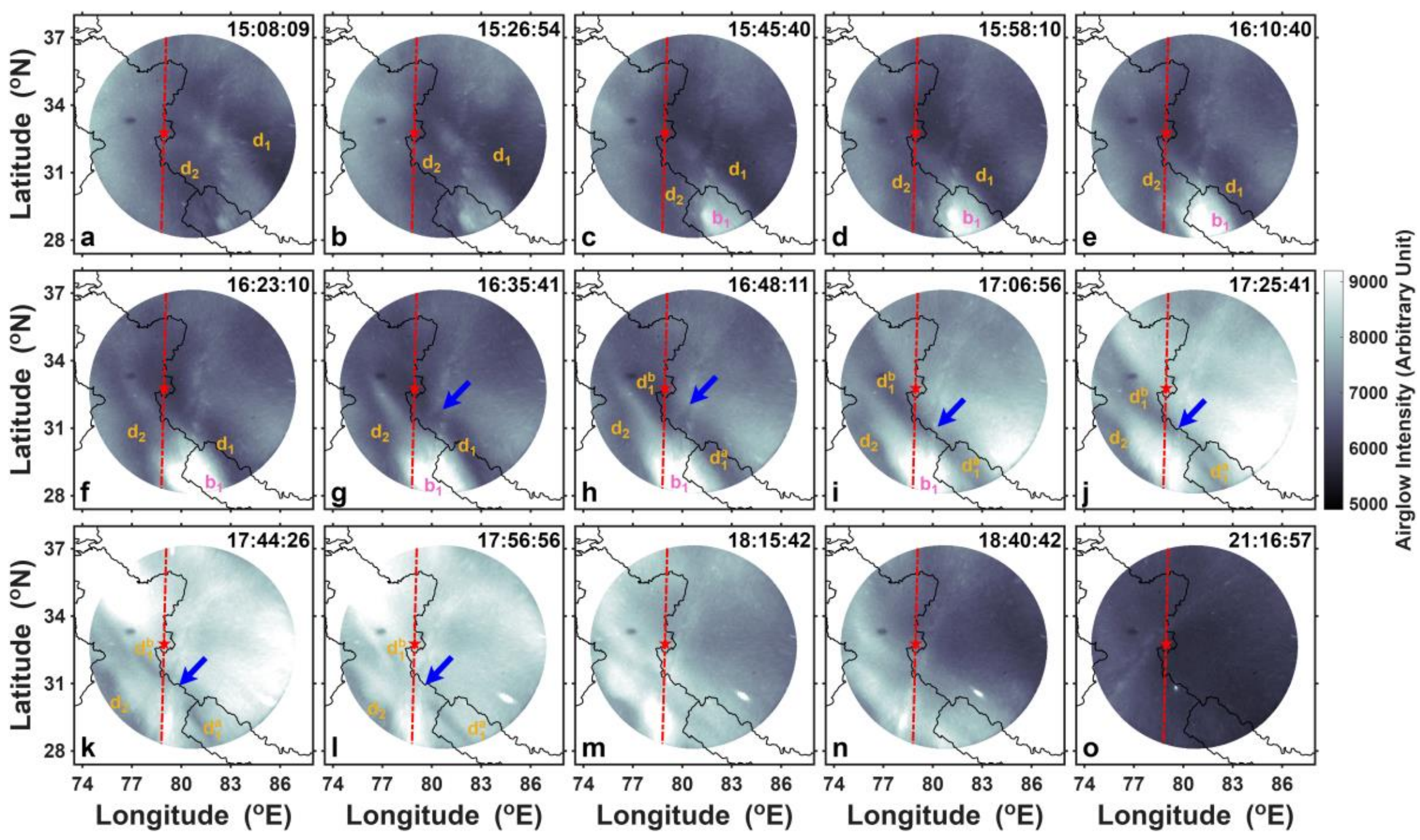

**Figure 1: (a-o)** sequence of O($^1$D) 630.0 nm airglow images. Red stars and dashed-dotted lines represent the location of the Hanle imager and geomagnetic field lines over Hanle, respectively. Two MSTID fronts and a plasma blob are denoted by '$d_1$', '$d_2$', and '$b_1$', respectively. The colored arrows in panels **(g-l)** indicate that '$d_1$' becomes distorted from its central portion, after which the band appears to bifurcate into '$d_1^a$' and '$d_1^b$', as shown in panels **(h-l)**.

### 3.2 Generation and evolution of plasma blob evident from global VTEC maps

Figures 2a-2o represent the 15 minutes cadence global VTEC maps provided by the IGS for 06 July 2021. In these VTEC maps, the colored stars and circles represent the location as well as the FOV of the Hanle imager, respectively. A localized enhanced VTEC region (hereafter referred to as the plasma blob) seemed to be generated beyond the southern edge of the imager's FOV around 15:30:00 UT, shown by a maroon colored ellipse in Figure 2b. After that, this region evolved and entered the imager's FOV from the southern edge around 15:45:00 UT (Figure 2c). As mentioned in the earlier section, the plasma blob entering the imager's FOV from the southern edge is also visible in the airglow images, as shown in Figure 1c. Similar to the observation in the airglow images, the plasma blob also propagated westward, as evident from the global VTEC maps, where it is denoted as '$b_1$' (Figures 2c-l). We also examined the global VTEC maps for the previous geomagnetically quiet night (05 July 2021) and found no localized region of enhanced VTEC indicative of a plasma blob. For the readers' convenience, a Movie 2 showing the sequence of global VTEC maps for 05 July 2021 and 06 July 2021 has been provided in the supplementary file.

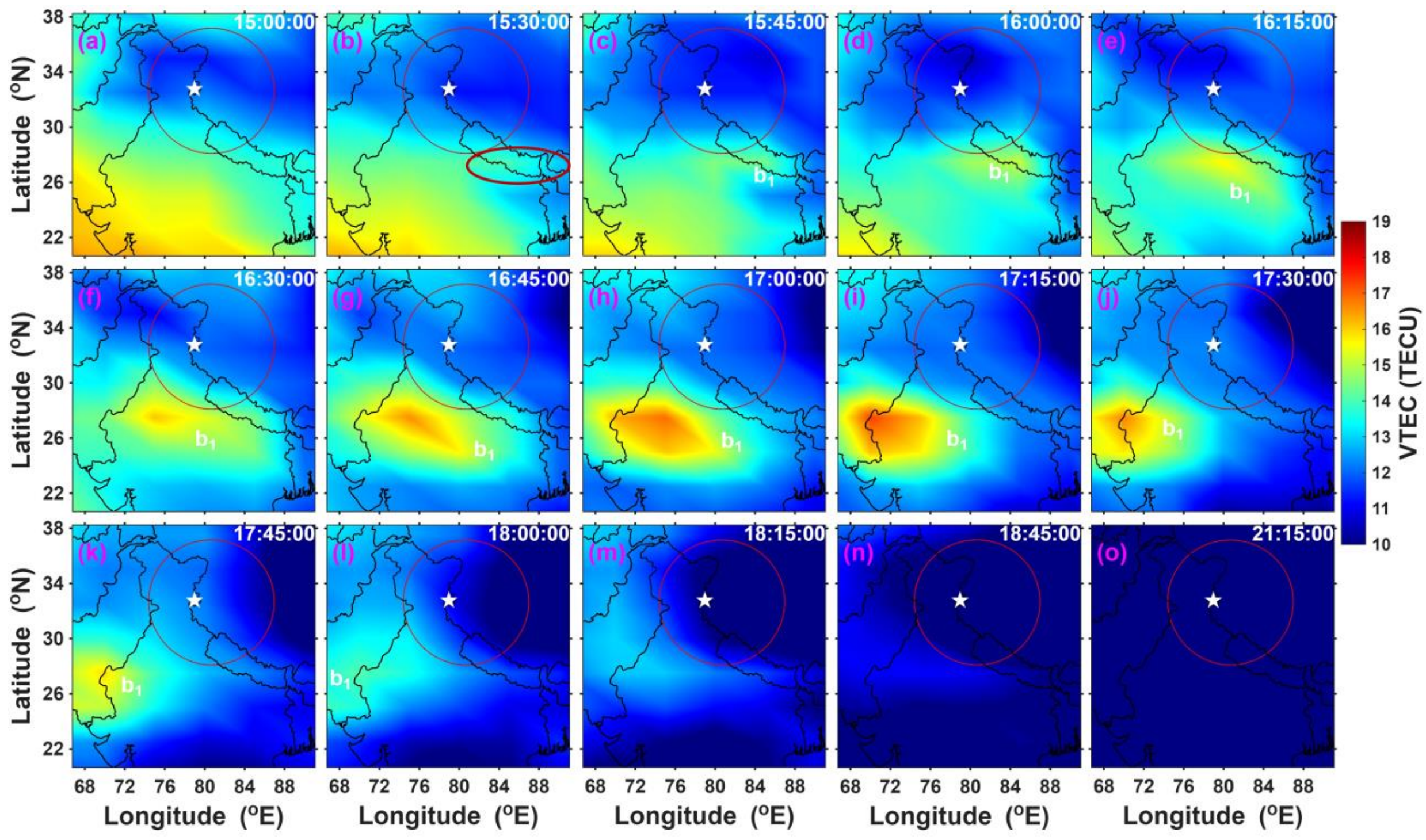

**Figure 2: (a-o)** sequence of IGS global VTEC maps. Colored stars and circles represent the location of the Hanle imager and its FOV, respectively. The colored ellipse in **(b)** represents the region of generation of the plasma blob (localized region of high VTEC). The plasma blob is later denoted by '$b_1$' in **(c-l)**.

### 3.3 Detection of MSTID band beyond the southern edge of the imager's FOV

In order to examine the associations, if any, between the variations in VTEC and the plasma blob and phase fronts of MSTID, we used VTEC data from multiple GNSS receivers recorded on 06 July 2021, as shown in Figure 3. The colored curves over the airglow images and global VTEC maps in Figures 3a-3l indicate the locations of ionospheric pierce points (IPPs) for different Pseudo-Random Noise (PRN) numbers recorded by multiple GNSS receivers. These include PRNs-32, 31, 10 and 31, 31, and 31 recorded from BAGD (88.7°E, 27.1°N), KUGE (85.5°E, 27.6°N), SNDL (85.7°E, 27.3°N), LUCK (80.6°E, 27°N), and GNFC (78°E, 30.4°N) GNSS receivers, respectively. The colored diamonds in these images represent the location of the GNSS receivers. Previously, several studies have demonstrated that detrending the total electron content (TEC) is effective for detecting fluctuations associated with MSTID structures (Ding et al., 2011; Otsuka et al., 2011; Patgiri et al., 2024c, 2025; Perwitasari et al., 2022; and references therein). For examining the temporal fluctuations in VTEC caused by plasma MSTID and plasma blob, the VTEC data were detrended by subtracting the background computed as a running mean over a ±30-minute window of the original VTEC series. The temporal fluctuations of the original VTEC series and its detrended series (DVTEC) are shown in Figures 3m-3r and 3s-3x, respectively. The color schemes for the PRN trajectories, as well as for the corresponding VTEC and DVTEC variations along each PRN, are kept consistent for the reader's convenience. The white and black circles in the PRN trajectories in Figures 3a-3l represent the locations of the IPPs corresponding to each PRN, in proximity to the airglow images and the global VTEC maps recording periods. The VTEC and DVTEC values for these times are also shown as white and black circles in Figures 3m-3x.

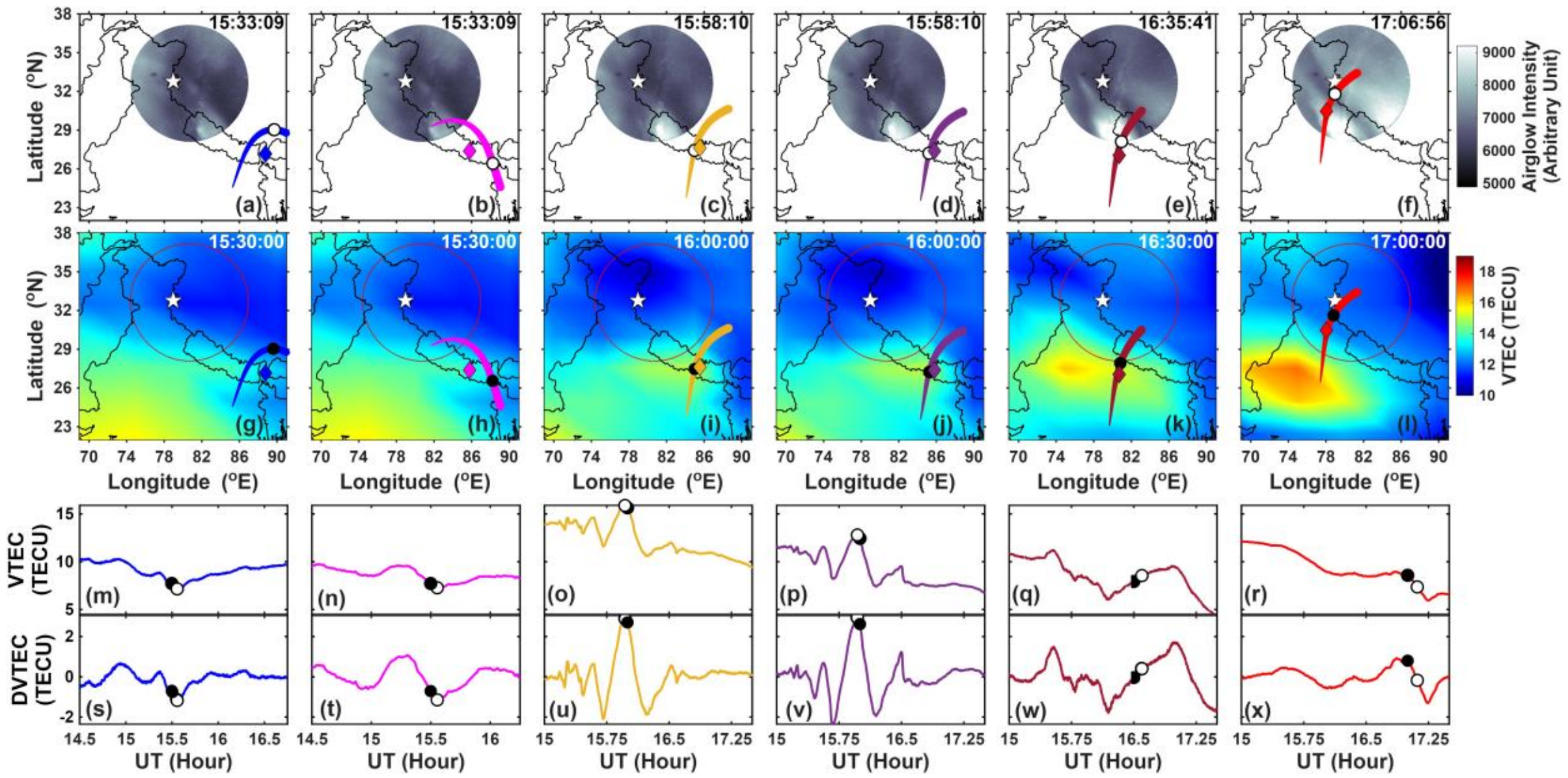

**Figure 3: (a-l)** The colored curves over the airglow images and global VTEC maps represent the IPP points for the PRNs-32, 10, 31, 31, 31, and 31 recorded from BAGD, SNDL, KUGE, SNDL, LUCK, and GNFC, receivers, respectively. The colored diamonds in these images represent the location of the GNSS receivers. **(m-r)** and **(s-x)** denote the temporal fluctuations of VTEC and DVTEC, respectively. The white and black circles in **(a-l)** represent the locations of IPP points for PRNs close to airglow and global VTEC observations, respectively. Around those times, the corresponding VTEC and DVTEC values are also denoted by white and black circles in **(m-x)**.

It is noteworthy that the VTEC along PRN-31, recorded from the GNFC receiver, showed a decrease when the PRN crossed one of the plasma-depleted (dark) fronts of the MSTID (Figures 3f and 3r). The corresponding DVTEC series showed variations of approximately 2 TECU (Figure 3x). This is consistent with previous studies that reported DVTEC fluctuations of approximately 0.5-2 TECU associated with MSTID activity (Ding et al., 2011; Huang et al., 2016; Patgiri et al., 2024c; Sun et al., 2020). In addition, both VTEC and DVTEC values along PRN-31 for the LUCK receiver showed significant enhancements when the PRN crossed the portion of the plasma blob region observed in both the airglow images and the global VTEC maps (Figures 3e, 3k, 3q, and 3w). In this case, an enhancement of nearly 3.5 TECU was observed in both VTEC and DVTEC. More interestingly, VTEC and DVTEC values along PRNs 32 and 10, recorded from the BAGD and SNDL stations, respectively, showed depletions as the PRNs crossed beyond the southern and southeastern edges of the FOV (Figures 3a-3b, 3g-3h, 3m-3n, and 3s-3t). This particular region appears to be the lower-latitude extension of the plasma-depleted front of MSTID (i.e., '$d_1$'). The maximum depletions of the VTEC/DVTEC values were observed near 29°N and 89.7°E from BAGD and 26.4°N and 88.2°E from SNDL. In a similar manner, we investigated the presence of the plasma blob beyond the southern boundary of the imager's FOV, which, however, was evident only in the global VTEC maps. It is clear from Figures 3c-3d and 3i-3j that PRNs-31, recorded from the KUGE and SNDL receivers, crossed the region of the plasma blob observed in the global VTEC maps but moved beyond the southern edge of the imager's FOV. Significant enhancements in VTEC and DVTEC were recorded from both receivers (Figures 3o-3p and 3u-3v). The observed enhancements were 4.3 TECU and 4.8 TECU in VTEC and DVTEC, respectively, at KUGE, and 4.5 TECU and 5.2 TECU at the SNDL receiver. Notably, the enhancement began around 27.0°N, 84.8°E/26.7°N, 85°E and reached a maximum near 27.5°N, 85°E/27.2°N, 85.2°E for the KUGE/SNDL receiver.

**3.4 In-situ Characteristics of Ion Density Variations within the Plasma Blob**

The preceding sections describe the spatiotemporal evolution of 2D characteristics of plasma blob using optical remote-sensing instrument, such as airglow imager, and radio remote-sensing technique, such as GNSS-derived VTEC. In addition to their detection in airglow images and VTEC measurements, previous studies have shown that plasma blobs also exhibit distinct signatures in in situ ion density measurements from Low Earth Orbit (LEO) satellites (Kil et al., 2011; Miller et al., 2014 Park et al., 2003). These characteristics include observed enhancements in $O^+$ density and depletions in $H^+$ density. In order to examine these features, we utilized the total ion density, $O^+$, and $H^+$ density measurements at around 545 km altitude by IVM onboard FORMOSAT-7/COSMIC-2 (Figure 4). The measurements obtained on 06 July 2021 were compared with those from the previous geomagnetically quiet night of 05 July 2021 (Ap = 7), when no evidence of plasma blobs was observed in the global VTEC maps. The curves overlaid on the global VTEC maps in Figures 4a-4b indicate the FORMOSAT-7/COSMIC-2 satellite trajectories for 05 July 2021 (green) and 06 July 2021 (red) obtained nearly coincident with the satellite measurements. The corresponding total ion, $O^+$, and $H^+$ densities are shown in Figures 4c-4e. The shaded area in Figures 4c-4e indicates the interval during which the satellite passed over the southernmost portion of the plasma blob, as evident from its trajectory over the global VTEC map on 06 July 2021 (Figure 4b). Interestingly, during this period, the measurements on 06 July 2021 showed enhancements of approximately 40% in total in-situ ion density and about 58% in $O^+$ density compared to those on 05 July 2021 (Figures 4c-4d). However, during this interval on 06 July 2021, the $H^+$ density showed a ~50% reduction compared to that on 05 July 2021 (Figure 4e).

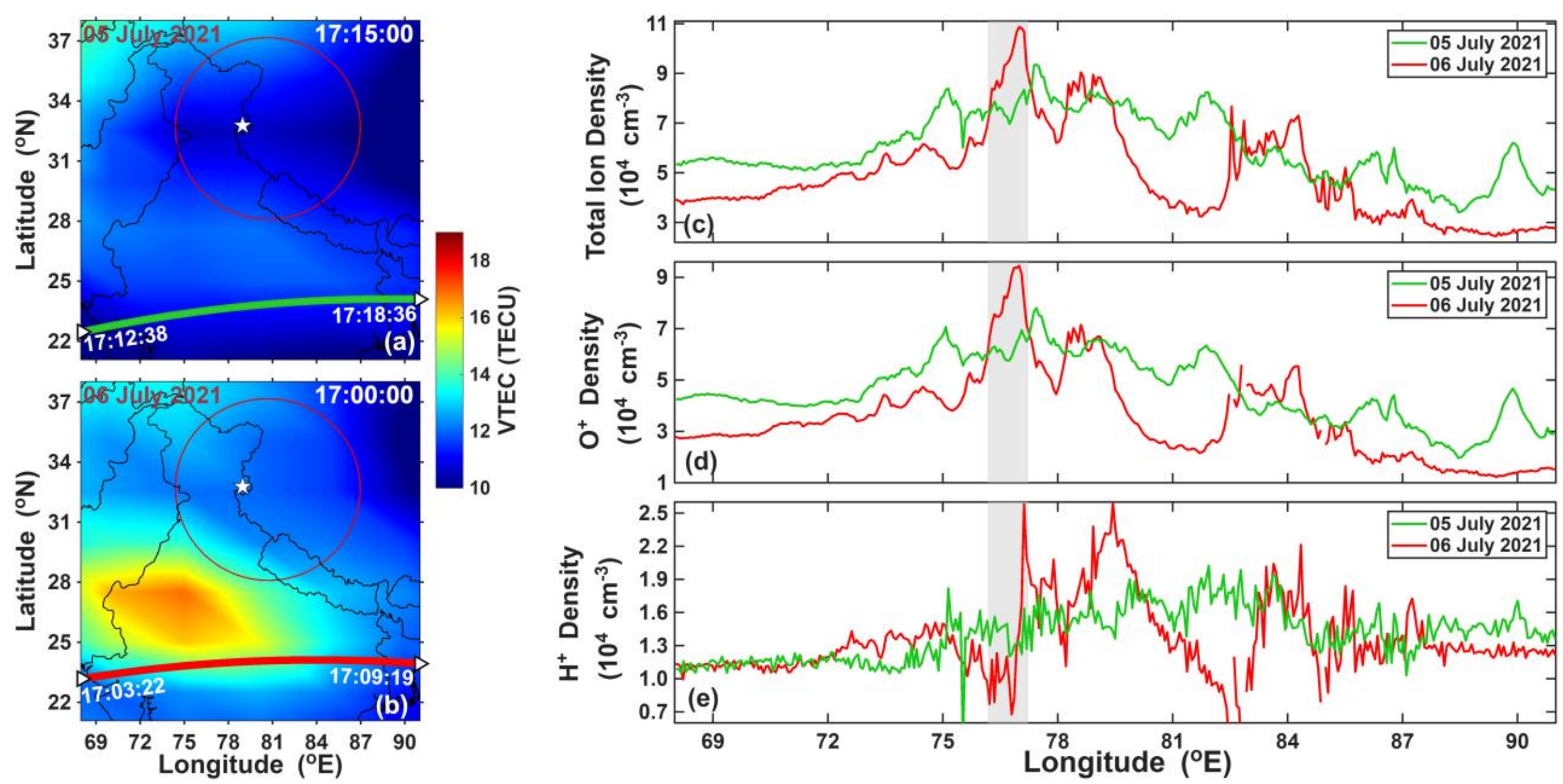

**Figure 4: (a-b)** The trajectories of FORMOSAT-7/COSMIC-2 on 05 July 2021 (green) and 06 July 2021 (red) are plotted over the corresponding global VTEC maps, recorded around the same time as the FORMOSAT-7/COSMIC-2 measurements. **(c-e)** Longitudinal variations of total ion, $O^+$, and $H^+$ densities for the two nights. The shaded area in panels **(c-e)** represents the segment of the FORMOSAT-7/COSMIC-2 path intersecting the plasma blob on 06 July 2021, as seen in the global VTEC map, where total ion and $O^+$ densities increased while $H^+$ density decreased relative to 05 July 2021.

### 3.5 Absence of low-latitude ionospheric irregularities evident from ROTI maps and ionosonde observations

Previous studies have demonstrated that the generation mechanism of plasma blobs is closely associated with the development of equatorial plasma bubbles (EPBs) that originate at lower latitudes and extend toward the plasma blob observation region (Anderson & Haerendel, 1979; Haaser et al., 2012; Krall et al., 2009, 2010). In order to examine the presence of any low-latitude origin ionospheric irregularities like EPBs/equatorial spread-F (ESF) on 06 July 2021, ROTI maps over equatorial regions and ionograms from low-latitude station Gadanki (13.5°N, 79.2°E; Mlat.

~6.5°N) were utilized (Figures 5a-5o and 5a$_1$-5o$_1$). Generally, ROTI magnitude ≥ 0.25 TECU/min is a possible indication of EPB/ESF occurrence (Ji et al., 2015; Nishioka et al., 2008). In ROTI maps, absence of any EPB/ESF was confirmed as ROTI values were less than 0.25 TECU/min before, during, and after the observations of plasma blob and MSTID on 06 July 2021 (Figures 5a-5o). Furthermore, ionograms recorded from a low-latitude station, Gadanki, did not show spread-F signatures prior to, during, and after the time of plasma blob and MSTID observations (Figures 5a$_1$-5o$_1$).

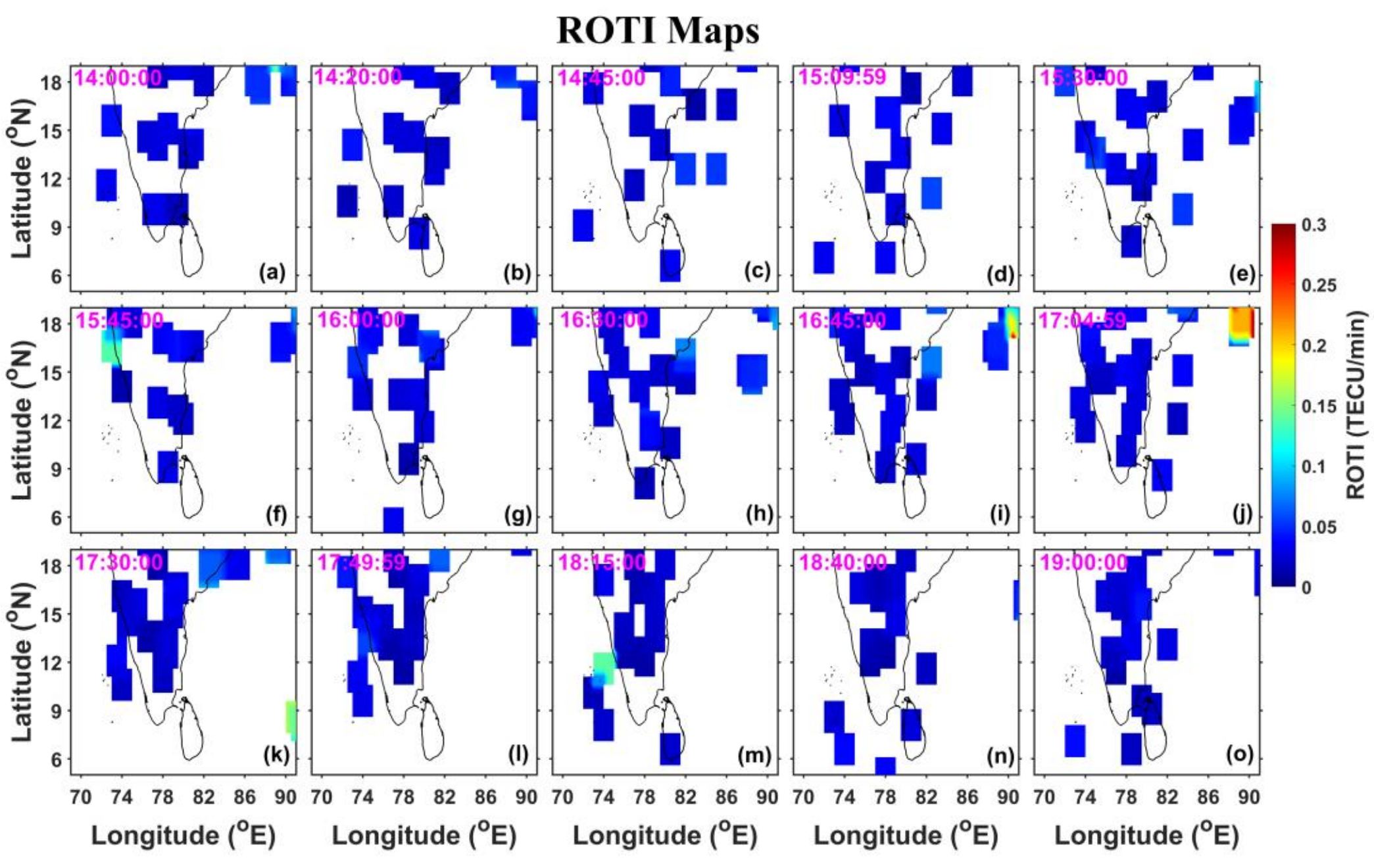

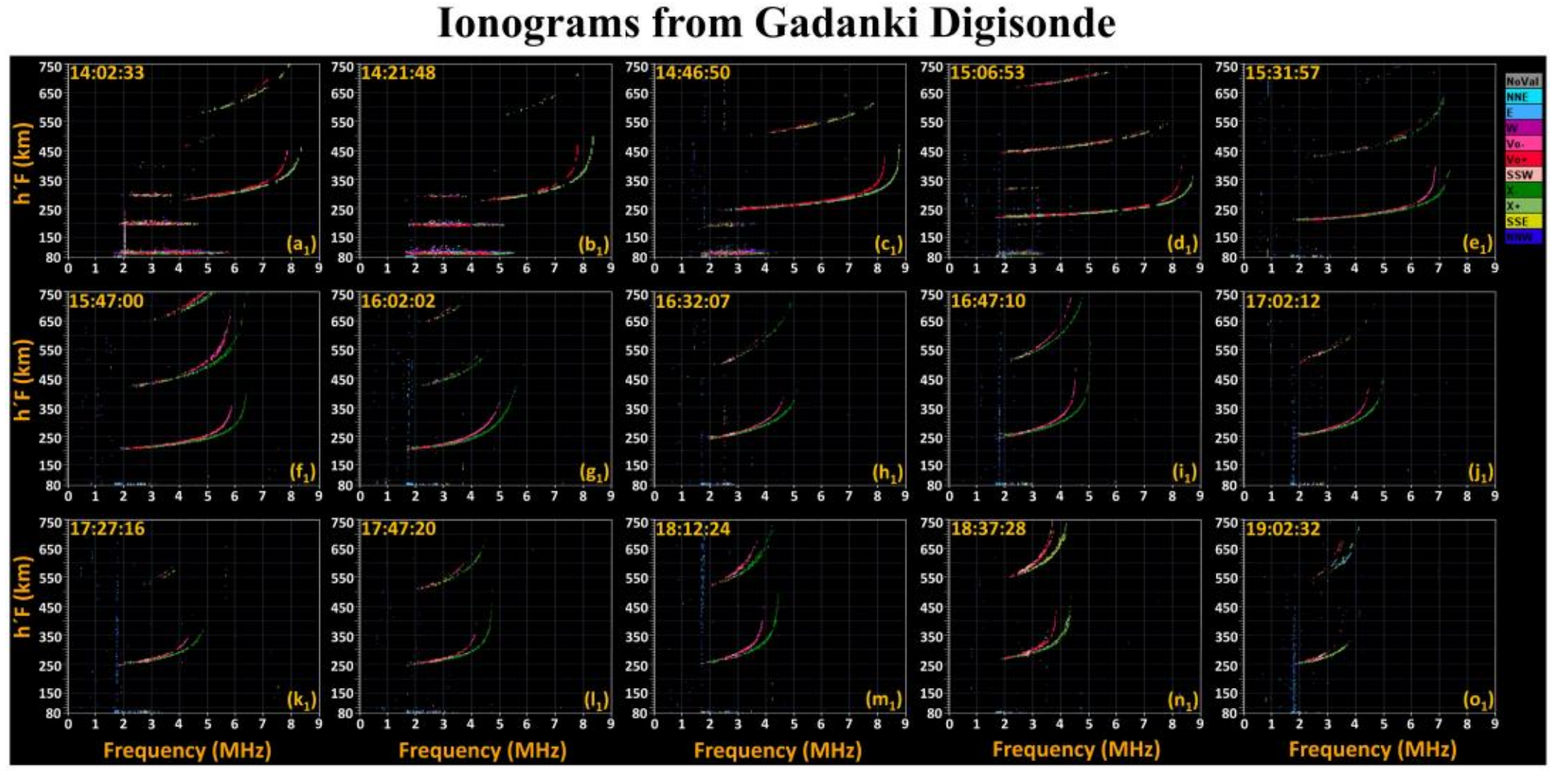

**Figure 5:** Sequence of **(a-o)** ROTI maps over low-latitude regions and **(a$_1$-o$_1$)** ionograms recorded by Gadanki digisonde on 06 July 2021.

### 3.6 Evidence of sporadic-E ($E_S$) layers near the observation region of MSTID and plasma blob

Extensive studies have demonstrated that the electrodynamic coupling between the E and F-regions can amplify the growth rate of Perkins instability, thereby facilitating the generation of MSTIDs (Cosgrove & Tsunoda, 2004; Otsuka et al., 2007; Yokoyama et al., 2009). Polarization electric fields from $E_S$-layer irregularities can map to the F-region, amplifying the Perkins instability and leading to the formation of MSTIDs. The presence of $E_S$-layers can be investigated through the variation of altitudinal profiles of zonal wind within the E-region, as these layers typically form near the nodes of vertical shear in zonal wind (Axford, 1963; Patgiri et al., 2024c; Whitehead, 1961; Yamazaki et al., 2022). The present study utilizes zonal wind profiles derived from green-line emission measurements recorded by ICON/MIGHTI to examine the possible occurrence of $E_S$-layers. The tangent points of these measurements, located near the region of MSTID observation prior to its appearance in airglow images on 06 July 2021, are plotted as colored curves over the global VTEC maps corresponding to nearly the same observation times. (Figure 6g). The zonal wind profiles of these measurements showed significant vertical shear as shown in Figure 6m. The statistical 1-sigma uncertainties in zonal wind estimates from the ICON/MIGHTI dataset are represented as horizontal error bars in the zonal wind profiles.

In addition, the presence of $E_S$-layers can be investigated using the SNR occultation profiles of the L1 band, as the strong vertical ion density gradients within $E_S$-layers significantly affect L1 signal propagation (Arras & Wickert, 2018; Hajj et al., 2002; Patgiri et al., 2024c; Tsai et al., 2018; Yamazaki et al., 2022). The colored curves over the airglow images and global VTEC maps in Figures 6d-6f and 6h-6l, respectively, represent the locations of tangent points corresponding to the L1-band SNR measurements by FORMOSAT-7/COSMIC-2 on 06 July 2021, taken near the observation times of the airglow images and global VTEC maps. It is to be noted that, subfigures **(a-c)** are left blank due to the unavailability of airglow images during those specific times. The combined measurements provide comprehensive coverage of the region surrounding the MSTID and plasma blob, capturing their evolution before, during, and after their detection in the airglow images and global VTEC maps. The normalized SNR profiles ($[SNR-SNR_0]/SNR_0$) of these measurements with same color scheme as the tangent locations are shown in Figures 6n-6r. For normalization, the background SNR ($SNR_0$) was obtained by applying a moving mean to the original SNR data using a 20 km sliding window. Finally, the moving standard deviation of the normalized SNR profiles was calculated using a 1 km sliding window, and these standard deviation profiles are plotted as blue curves in Figures 6n-6r. A similar approach was previously employed by Arras and Wickert (2018), Tsai et al. (2018), and Yamazaki et al. (2022) to investigate the occurrence of $E_S$-layers. Following the criteria outlined in previous studies, an $E_S$-layer is identified when the following conditions are met: **(a)** the standard deviation of the normalized SNR profile exceeds a threshold of 0.2, **(b)** the standard deviation near the top of the occultation remains below 0.2, and **(c)** the altitude corresponding to the peak standard deviation is higher than 80 km to avoid potential contamination from water vapor effects on the L1-band SNR. All three criteria are satisfied in the five profiles shown in Figures 6n-6r. This confirms the presence of $E_S$-layers near the region (~27°-37°N, ~69°-90°E) where MSTID and plasma blob were observed in the airglow images and global VTEC maps as shown in Figures 1 and 2.

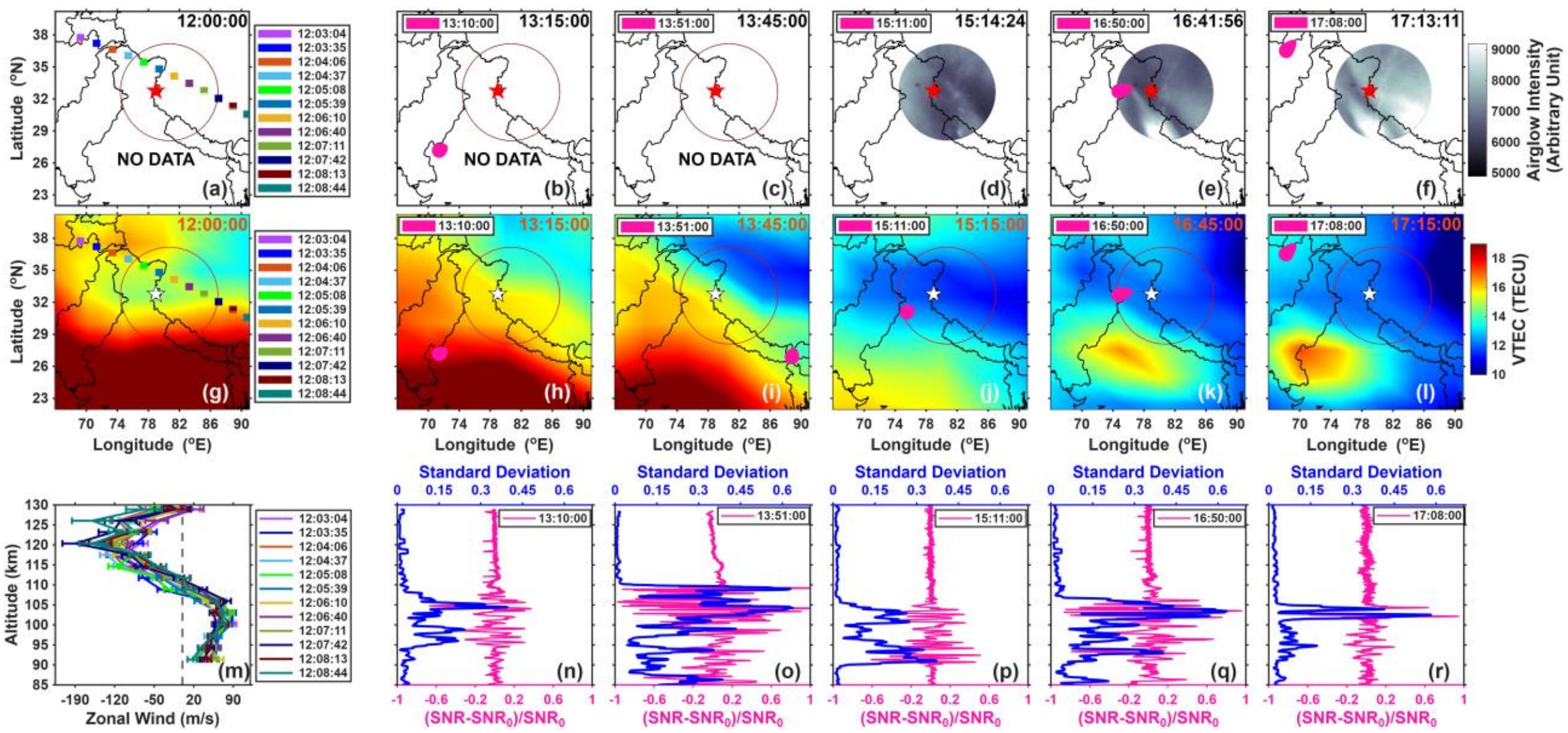

**Figure 6: (g)** The colored lines over the global VTEC maps illustrate the tangent points of the wind observations (green line) from ICON/MIGHTI during the global VTEC observation period at 12:00 UT, shown in the top-right corners. **(m)** The corresponding altitudinal profiles of zonal wind measured by ICON/MIGHTI for all the scans, with 1-sigma statistical uncertainties represented by error bars. The magenta curves over the airglow images **(d-f)** and global VTEC maps **(h-l)** indicate the locations of five FORMOSAT-7/COSMIC-2 SNR profile measurements taken near the times of global VTEC and airglow observations shown in the top-right corners. **(n-r)** The magenta and blue curves denote the normalized FORMOSAT-7/COSMIC-2 RO profiles of SNR and their standard deviation, respectively, obtained using a 1 km sliding window. Subfigures **(a-c)** are left blank due to the unavailability of airglow images during those specific times.

## 4. Discussion

As discussed in the previous section, multi-instrument analyses revealed simultaneous detection of two different types of ionospheric perturbation structures. These observations indicate the concurrent presence of an MSTID and a localized plasma density enhancement, commonly referred to as a plasma blob, during the geomagnetically quiet night of 06 July 2021. Both the MSTID structure consisting of plasma depleted (dark) fronts ('$d_1$', '$d_2$', '$d_1^a$', '$d_1^b$') and plasma blob ('$b_1$') were observed in O($^1$D) 630.0 nm airglow images (Figure 1). The plasma blob initially appeared as a localized VTEC enhancement near 27.5°N at 15:30 UT (ellipse, Figure 2b) before evolving into the imager's FOV nearly around 15:45 UT (Figures 1c and 2c). An intriguing feature evident in the airglow images in Figure 1 is the interaction between the plasma blob and a dark front of MSTID (i.e., between '$b_1$' and '$d_1$'). Following the interaction, the central portion of '$d_1$' (indicated by arrows in Figures 1g-1l) gradually became distorted, eventually causing '$d_1$' to appear visually bifurcated into two distinct dark fronts, designated '$d_1^a$' and '$d_1^b$' (Figures 1h-1l). Additionally, the lower-latitude extent of '$d_1$' was estimated to be near 26.4°N, 88.2°E, based on the VTEC and DVTEC fluctuations observed along the PRN-10 satellite-receiver path at the SNDL station (Figures 3b, 3h, 3n, 3t). Interestingly, the corresponding peak VTEC/DVTEC enhancements for PRN-31 associated with the plasma blob occurred ~27.5°N, 85°E and 27.2°N, 85.2°E at the KUGE and SNDL stations, respectively (Figures 3c-3d, 3i-3j, 3o-3p, 3u-3v). These observations suggest that the lower-latitude extension of '$d_1$' was most likely located around the similar latitude of the plasma blob.

Apart from the localized regions of enhanced airglow intensity and elevated VTEC values, multiple previous studies reported that satellite measurement of in-situ plasma density associated with the plasma blob can exceed 40-60% above background plasma density (Kil et al., 2019; Pimenta et al., 2004; Rather et al., 2024). In this study, FORMOSAT-7/COSMIC-2 IVM measurements show a ~40% enhancement in total ion density within the blob (Figure 4c). Correspondingly, $O^+$ density enhanced by ~58% and $H^+$ density depleted by ~50% relative to the background (Figures 4d-4e). Such LEO-altitude (here ~545 km) ion composition transitions within plasma blobs are well-documented (Kil et al., 2011; Miller et al., 2014; Park et al., 2003). This feature has been schematically illustrated in Figure 4 of Miller et al. (2014). F-layer ascent aligns the $O^+$ density peak with the satellite altitude, resulting in $O^+$/$H^+$ enhancements/depletions, even though $H^+$ remains the dominant ion species at higher altitudes. This type of uplift observed within the plasma blob is in clear contrast to the behavior reported within the plasma-enhanced phase front of nighttime MSTIDs. The typically northwest-southeast (NW-SE) aligned plasma-enhanced front of nighttime MSTIDs is associated with southwestward-directed polarization electric fields, whose westward component facilitates a downward $\boldsymbol{E} \times \boldsymbol{B}$ drift (Narayanan et al., 2018; Otsuka et al., 2007; Patgiri et al., 2024a; Shiokawa et al., 2003b; Tsunoda. 2006; Yokoyama et al., 2009). Therefore, the localized plasma-enhanced region (plasma blob) observed in the present study exhibited characteristics that differ from those of the plasma-enhanced fronts associated with MSTIDs, indicating that it was not part of an MSTID structure.

As discussed in the previous section, the region containing the plasma blob was associated with an upward motion of the topside F layer, as indicated by the in-situ plasma density measurements from FORMOSAT-7/COSMIC-2. It is to be noted that the upliftment of F-layer typically lowers the charge transfer rate between $O^+$ and $O_2$, reducing dissociative recombination (Patgiri et al., 2024b; Peterson & Vanzandt, 1969; Takahashi et al., 1990). This results in the localized depletion of $O(^1D)$ 630.0 nm airglow emission. Hence, if the airglow imager had covered the plasma blob generation region (~27.5°N) seen in the global VTEC maps, situated just beyond the southern edge of its FOV, it would have recorded a localized depletion rather than enhanced airglow intensity. Instead, as evident from the global VTEC maps, the evolving blob entered the imager's FOV around 15:45:00 UT, appearing as a localized airglow enhancement that propagated westward (Figures 1c-1i, 2c-2l). The uniqueness and complexity of this scenario lie in a generation mechanism intricately linked to the MSTID, distinguishing it from earlier reports.

It is well established that the generation mechanism of plasma blobs is conventionally associated with that of equatorial plasma bubbles (EPBs) (Anderson & Haerendel, 1979; Haaser et al., 2012; Krall et al., 2009, 2010). The concept of a bubble-blob connection primarily stems from observations showing the occurrence of both plasma bubbles and blobs along the same magnetic meridian as well as notable similarities in their spatial distributions (Huang et al., 2014; Le et al., 2003; Martinis et al., 2009; Park et al., 2003, 2008; Yokoyama et al., 2007). However, the ROTI maps presented in Figures 5a-5o show no evidence of EPB activity, with values remaining well below the 0.5 TECU/min threshold typically associated with EPB/ESF occurrences (Ji et al., 2015; Nishioka et al., 2008). This condition persists across the lower-latitude sector of the plasma blob observation region during both its formation and subsequent evolution (Figures 5a-5o). Additionally, no spread-F signatures were observed in ionograms recorded by an ionosonde located at the low-latitude station Gadanki during the generation and evolution of the plasma blob on 06 July 2021 (Figures $5a_1$-$5o_1$). Collectively, these results indicate that the plasma blob observed on 06 July 2021 was not associated with the low-latitude EPB/ESF activity. While plasma blobs are frequently observed in tandem with EPBs, observational evidence demonstrates that they can also develop independently (Choi et al., 2012; Haaser et al., 2012; Kil et al., 2011). Furthermore, a limited number of studies have also documented the generation of plasma blobs in association with MSTIDs, inferred from their similar characteristics and simultaneous observations (Kil and Paxton, 2017; Kill et al., 2019; Miller et al., 2014). However, a confirmed theoretical explanation has not yet emerged, and it remains an open area of research. In this study, O($^1$D) 630.0 nm all-sky airglow images reveal the simultaneous occurrence of an MSTID and a plasma blob during the geomagnetically quiet night of 06 July 2021 (Figures 1a-1i). Given the total absence of lower-latitude EPB/ESF signatures, we propose a distinct electrodynamical generation mechanism linked to the MSTID, which is detailed in the following section.

In order to delineate the generation mechanism of the plasma blob observed on 06 July 2021, a schematic diagram is presented in Figure 7. In Figures 7a and 7b, three colored magnetic field lines intersect the MSTID structures in the airglow image near the peak 630.0 nm emission altitude (~250 km; cross symbols) and the initial blob formation region (ellipse) in the global VTEC map (asterisk symbols) across varying altitudes. An MSTID dark front is projected south of the imager's FOV in Figure 7a based on observed VTEC/DVTEC fluctuations (Figures 3m-3n, 3s-3t). Figure 7c illustrates the altitude vs latitude variations of these field lines, which cross the MSTID front near the peak 630.0 nm emission altitude (cross symbols/dashed line) and intersect the plasma blob generation area at different altitudes (asterisk-marked segments/dash-dotted lines). Figure 7d provides a 3D altitude-latitude-longitude representation focusing on the magenta field line intersecting the initial blob generation region at ~332 km across three horizontal reference planes: the $E_S$-layer-layer (~100 km), the 630.0 nm emission peak (~250 km), and the blob base region (~332 km). Notably, the $E_S$-layer is assumed to be located at ~100 km although it is extremely challenging to locate the exact altitude of layer. Nonetheless, since zonal wind shear node is located near ~110 km, with significant variations in SNR RO profiles of FORMOSAT-7/COSMIC-2 observed between 90 and 110 km, considering the $E_S$-layer to be located ~100 km is reasonable. Horizontal arrows in Figure 7d denote the polarization electric field ($\boldsymbol{E_P}$) mapped along magnetic field lines from either $E_S$-layer irregularities or MSTID dark fronts. $\boldsymbol{E_P}$ is assumed to be northeastward, consistent with the polarization fields observed within plasma-depleted MSTID fronts (Otsuka et al., 2007; Shiokawa et al., 2003b). This mapping is electrodynamically feasible because $E_S$-layer polarization fields can map to the F-region to enhance Perkins instability growth and generate MSTIDs (Kelley & Makela, 2001; Otsuka et al., 2007; Patgiri et al., 2024c). A similar link between the generation of plasma blobs and polarization electric fields associated with $E_S$-layer has been discussed by Su et al. (2022). The black curves in Figures 7e and 7f display the FORMOSAT-7/COSMIC-2 electron density occultation profiles and their corresponding tangent point tracks. The vertical arrows in Figures 7d and 7e denote the vertical $\boldsymbol{E_P}\times\boldsymbol{B}$ drift driven by $\boldsymbol{E_P}$ from the initial or base region of plasma blob formation.

Crucially, the mechanism described in this study relies on the magenta colored magnetic field line, which intersects the initial blob formation zone at the F-layer peak altitude (~332 km) (Figures 7a-7c). For this field lines, $\boldsymbol{E_P}\times\boldsymbol{B}$ drift driven by $\boldsymbol{E_P}$, mapped either from $E_S$-layer or MSTID dark front, facilitated the vertical transport of plasma from around the F-region peak to higher altitudes through upliftment of the layer (Figures 7d-7e). Additionally, the ionosphere maintains quasi-neutrality, and the F-region (200-700 km) is dominated by $O^+$ ions that closely follow electron density in both variation and magnitude with altitude (Kelley, 2009). Consequently, uplift of the F-layer can produce enhanced total ion and $O^+$ densities at LEO satellite altitudes above the ambient F-peak. The blue ellipse in Figure 7e highlights a region of lower electron density relative to the uplifted plasma originating near the F-peak. Interestingly, FORMOSAT-7/COSMIC-2 observations show significant enhancements in total ion and $O^+$ densities at ~545 km on 06 July 2021 compared to 05 July 2021, when no plasma blob signatures were observed in the global VTEC maps (Figures 4a-4c). At the same time, the upward drift raised the F-layer to higher altitudes, reducing $H^+$ density at the satellite altitude (Figures 4d-4e) despite its increasing abundance with height (Kelley, 2009). Similar $O^+$/$H^+$ density variations observed from LEO satellites during plasma blob events have been reported in lietratures (e.g. Kil et al., 2011; Miller et al., 2014; Park et al., 2003). However, to the best of our knowledge, the role of polarization electric field of MSTID dark front in producing the observed $O^+$/$H^+$ density variations associated plasma blob events has not been explicitly reported. Furthermore, the nighttime uplift of the F-layer helped in sustaining TEC for a longer duration over this region than over surrounding areas on 06 July 2021, since chemical losses were lower at higher altitudes (Bailey et al., 1992; Balan et al., 1994). This resulted in a localized TEC enhancement that appeared as a plasma blob, as evidenced by the global VTEC maps and the VTEC fluctuations recorded by multiple GNSS receivers on 06 July 2021 (Figures 2 and 3). It should be noted that this interpretation is not applicable to the other two field lines, which intersect the initial plasma blob formation region well above the F-peak (~558 km and ~782 km), where plasma densities are substantially lower.

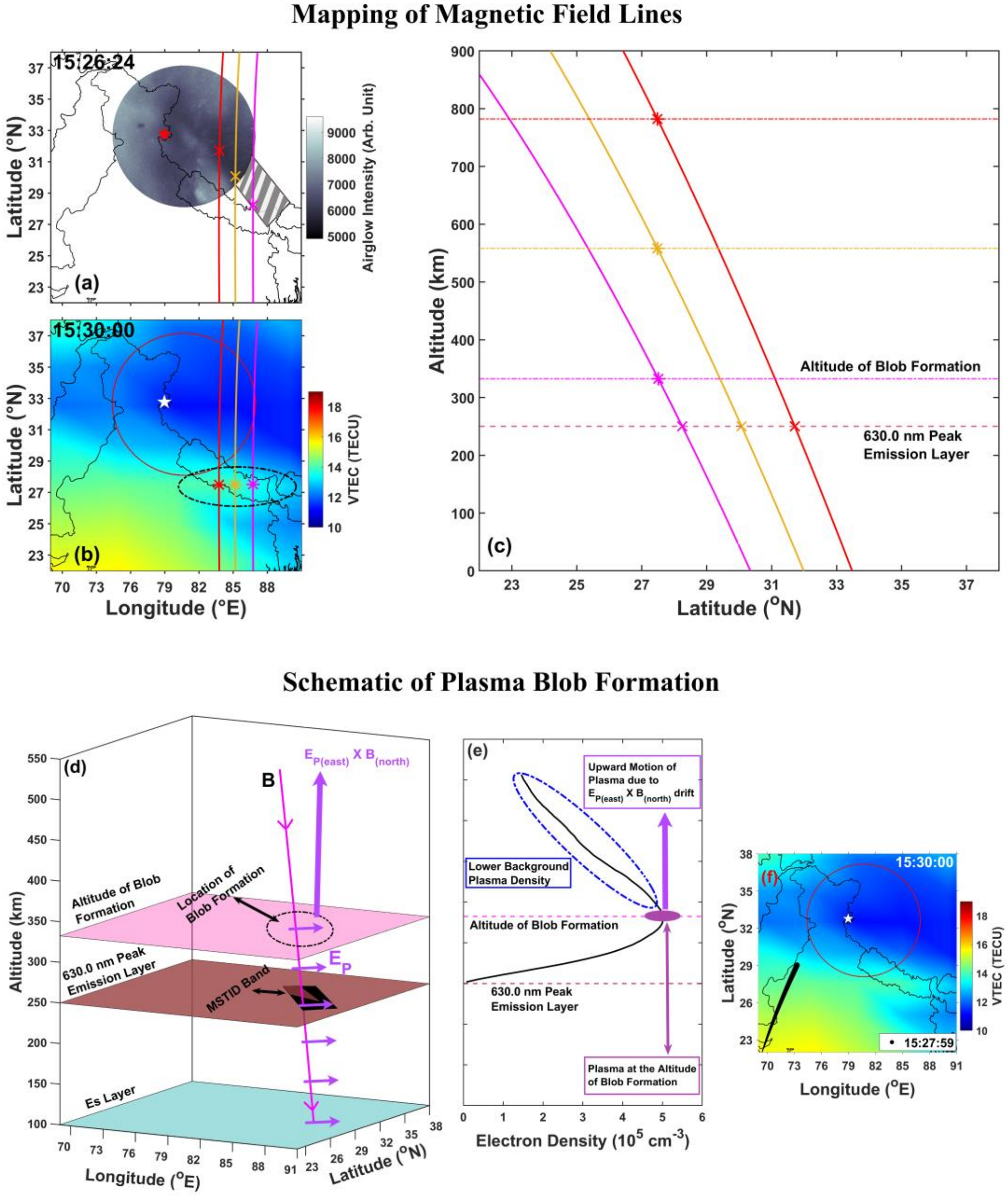

**Figure 7: (a, b)** Three colored lines overlaid on the global VTEC map and airglow image represent magnetic field lines. Cross markers in **(a)** indicate locations mapping onto the MSTID dark front near the 630.0 nm emission peak altitude, with the quadrilateral highlighting its lower-latitude extension, while asterisks in **(b)** mark intersections with the plasma blob region (black ellipse). Different horizontal lines denote the altitudes of the 630.0 nm peak emission layer and the altitudes where magnetic field lines cross the plasma blob. **(c)** Altitude variation of the three magnetic field lines as a function of latitude, with corresponding cross and asterisk markers. **(d)** Schematic showing three altitude planes corresponding to the *E*s-layer, 630.0 nm emission peak, and plasma blob formation height. The MSTID dark front, plasma blob formation region, and intersecting field line are illustrated by a black rectangle, an ellipse, and a magenta curve, respectively. **(e, f)** Electron density profiles from FORMOSAT-7/COSMIC-2 and corresponding tangent points, with horizontal lines in **(e)** indicating the 630.0 nm peak emission altitude and altitude of plasma blob generation. The vertical arrows in **(d)** and **(e)** denote the vertical $\boldsymbol{E_P}\times\boldsymbol{B}$ drifts from the location of plasma blob formation driven by the polarization electric field ($\mathbf{E_P}$) represented with horizontal arrows.

In addition to the enhancements in TEC and in-situ ion densities, an increase in $O(^1D)$ 630.0 nm airglow intensity was observed (Figures 1c-1i). At first glance, this appears inconsistent with the F-layer uplift, which is indirectly supported by the $O^+/H^+$ density variations (Figures 4c-4e). During nighttime, uplift (downlift) of the F layer decreases (increases) the charge-transfer reaction from $O^+$ to $O_2$ owing to lower (higher) $O_2$ densities at higher (lower) altitudes, resulting in reduced (enhanced) $O(^1D)$ 630.0 nm emission (Bittencourt & Sahai, 1979; Patgiri et al., 2024b; Rajesh et al., 2007). Notably, the initial (base) region of plasma blob formation evident from global VTEC map was located beyond the southern edge of the imager's FOV around 15:30 UT (Figure 2b) and therefore was not captured in the airglow images. After 15:30 UT, the enhanced VTEC region expanded poleward, entered the FOV around 15:45 UT, and subsequently propagated westward as a localized region of enhanced airglow intensity (Figures 1c-1i, 2b-2l). The blob's existence outside the FOV is further supported by GNSS-derived VTEC/DVTEC fluctuations and in-situ ion density measurements (Figures 3o-3p, 3u-3v, and 4c-4e). The observed airglow enhancement can be explained by the latitudinal evolution of the plasma blob. The polarization electric field ($\mathbf{E_P}$) associated with the MSTID dark front mapped to lower latitudes near the F-peak, uplifting high-density plasma that subsequently diffused along magnetic field lines to higher latitudes and lower altitudes (~250 km). Since the dissociative recombination rate depends on the F-region plasma density (Patgiri et al., 2024b; Rajesh et al., 2007; Taori et al., 2003), the influx of high-density plasma enhanced this reaction, resulting in an increase in the $O(^1D)$ 630.0 nm emission. Additionally, even though the global VTEC maps suggest substantial longitudinal expansion of the plasma blob (Figures 2b-2i), a similar feature is absent in the airglow images (Figures 1c-1i). This longitudinal growth may be apparent arising from the relatively coarse longitudinal resolution (5°) and interpolation applied to the longitudinal grid of the VTEC maps (Schaer et al., 1998). Therefore, in the absence of complementary datasets, it is difficult to adequately explain such a longitudinal feature of the plasma blob within the scope of the present study.

The preceding discussion has focused on plasma blob generation by the polarization electric field associated with the MSTID phase front. Besides electric fields, previous studies have shown that thermospheric neutral wind can also produce localized plasma density enhancements and plasma blobs. The combined effects of meridional winds and reversals of zonal wind can transform EPBs/ESFs into plasma blobs (Krall et al., 2010; Martinis et al., 2009), while the latitudinal convergence of meridional winds can also generate plasma blobs independently of plasma bubbles (Wu et al., 2022). Neither mechanism appears applicable in the present case. This is because no evidence of plasma bubbles is observed (Figures 1 and 5). Moreover, HWM and TIEGCM models indicate that the zonal wind remained predominantly eastward before, during, and after the event on 06 July and on the control night (05 July) at ~250 km and ~350 km, showing no sudden reversals (Movie 3). Furthermore, unlike the scenario reported by Wu et al. (2022), no significant latitudinal convergence of the meridional wind was present prior to or during the plasma blob event. These results therefore suggest that the observed plasma blob was primarily driven by polarization electric fields associated with the $E_S$-layers or MSTID phase front rather than by thermospheric neutral winds.

Beyond the generation of the plasma blob, a notable feature of this event observed in the airglow images was the apparent interaction between the plasma blob ('$b_1$') and an MSTID dark front ('$d_1$'), resulting in the front's gradual distortion and bifurcation (Figure 1). After appearing around 15:45 UT, '$b_1$' propagated westward together with '$d_1$', while localized brightening progressively developed within '$d_1$' (Figures 1c-1l). This brightening gradually distorted and thinned the central portion of '$d_1$' which is depicted by colored arrows in Figures 1g-1l. Additionally, the east-west extents of the northern and southern segments adjacent to the central portion of '$d_1$' appeared to gradually distort and become thinner between 16:35:41 UT and 17:56:56 UT (Figures 1g-1l). By 17:25:41 UT, the brightened region had completely separated the two segments ('$d_1^a$' and '$d_1^b$'), resulting in a clear bifurcation of the original dark front (Figures 1k-1l). This bifurcation was likely driven by the influx of plasma from the blob into the depleted region of '$d_1$', leading to its gradual filling. The elevated plasma density within the blob, relative to the depleted region inside the MSTID dark front, establishes a plasma pressure gradient that drives plasma flow from the blob toward the MSTID dark front.

## 5. Conclusion

Using a combination of ground and satellite-based observations, this study reports the generation and evolution of a plasma blob on the geomagnetically quiet night of 06 July 2021, likely driven by polarization electric field associated with MSTID phase front. The following conclusions are drawn from the various observational results:

**a)** The polarization electric field within the plasma-depleted MSTID front mapped to lower latitudes along magnetic field lines, facilitating the upward transport of plasma from the F-region peak to higher altitudes through the uplift of the ionospheric layer.

**b)** The electric fields associated with the underlying sporadic-E layers, which might facilitate the generation of MSTID by enhancing the Perkins instability growth rate, as reported in numerous previous studies. Therefore, in addition to the polarization electric field associated with the MSTID, electric fields generated in the sporadic-E layers might also have contributed to the ascend of the F-layer.

**c)** The upward transport of plasma led to a localized enhancement in TEC relative to the background, manifesting as a plasma blob, as plasma can persist for longer durations at higher altitudes than in the surrounding regions. Such upward transport to higher altitudes resulted in an enhancement of total and $O^+$ ion densities, accompanied by a reduction in $H^+$ density at LEO satellite altitudes.

**d)** The upward-transported plasma subsequently diffused along magnetic field lines and entered the imager's field of view near the $O(^1D)$ 630.0 nm emission layer altitudes. This led to a localized enhancement of the $O(^1D)$ 630.0 nm emission intensity, resulting from increased dissociative recombination of $O_2^+$, which manifested as a plasma blob in the airglow images.

**e)** An intriguing feature of the observed event was the interaction between the plasma blob and one of the MSTID phase fronts, which led to its gradual bifurcation into two fronts followed by the subsequent decay of the front. The influx of plasma from the blob into the MSTID front, driven by the density gradient, facilitated the bifurcation and subsequent decay of the front.

**Acknowledgments**

D. Patgiri acknowledges the fellowship from the Ministry of Education, Government of India for carrying out this research work. S. Sarkhel acknowledges the financial support from the Science and Engineering Research Board, Department of Science and Technology, Government of India (CRG/2021/002052) to maintain the multi-wavelength airglow imager. The support from Indian Astronomical Observatory (operated by Indian Institute of Astrophysics, Bengaluru, India), Hanle, Ladakh, India for the day-to-day operation of the imager is duly acknowledged. R. Rathi acknowledges the fellowship from NERC grant NE/W003090/1, School of Physics and Astronomy, Lancaster University, United Kingdom. This work is also supported by the Department of Space and the Ministry of Education, Government of India.

**Data Availability Statement**

The datasets analyzed in this study are available in Patgiri (2026). Global TEC maps from the International GNSS Service (IGS) are available through https://cdaweb.gsfc.nasa.gov/index.html. The ICON/MIGHTI Level 2.2 (Version 5) cardinal wind data are publicly available at https://cdaweb.gsfc.nasa.gov/pub/data/icon/. The FORMOSAT-7/COSMIC-2 Level 1b SNR RO profiles (caL1Snr) and IVM data are available through Taiwan Analysis Center for COSMIC: https://tacc.cwa.gov.tw/v2/trops_download.html. The GNSS observation and navigation files in RINEX format for the DNSG, KUGE, and SNDL receivers were obtained from the UNAVCO data archive (https://www.unavco.org/). Geomagnetic activity indices were obtained from the WDC, Kyoto (http://wdc.kugi.kyoto-u.ac.jp/kp/index.html). The ROTI data is available in https://stdb2.isee.nagoya-u.ac.jp/GPS/GPS-TEC/GLOBAL/.